\documentclass[aps,prx,reprint,superscriptaddress,footnote, twocolumn,notitlepage]{revtex4-1}

\usepackage{tikz}  
\usetikzlibrary{arrows,shapes,positioning,shadows,backgrounds,fit}

\usepackage{float}
\usepackage{relsize}
\usepackage{graphicx}
\usepackage{amsmath, bbm}
\usepackage{amssymb}
\usepackage{amsthm}
\usepackage{comment}
\usepackage{mathrsfs}
\usepackage{xcolor}
\usepackage{cancel}
\usepackage{bbold}
\usepackage{titlesec}
\usepackage{enumitem}  
\usepackage{stmaryrd}

\usepackage[normalem]{ulem}

\titlespacing{\section}{0ex}{2ex}{0.4ex}
\titleformat{name=\section}{\bf}{\thesection.}{.3em}{}

\usepackage{soul}
\usepackage{dsfont}
\usepackage{times}

\newcommand{\Wdiss}{W_{\text{diss}}}

\def\be{\begin{eqnarray}}
\def\ee{\end{eqnarray}}
\newcommand{\tr}[1]{\text{Tr}\left(#1\right)}
\newcommand{\<}{\langle}
\renewcommand{\>}{\rangle}
\newcommand{\ket}[1]{|{#1}\rangle}
\newcommand{\bra}[1]{\langle{#1}|}

\newcommand{\e}{{\mathcal{E}}}

\theoremstyle{plain}

\definecolor{myblue}{rgb}{0.2,0.2,0.8}
\definecolor{myblack}{rgb}{0,0,0}
\definecolor{myurl}{rgb}{0.1,0.1,0.4}

\usepackage[colorlinks=true,citecolor=myblue,linkcolor=myblack,urlcolor=myurl]{hyperref}

\begin{document}

\title{Generalised linear response theory for the full quantum work statistics} 

\author{Giacomo Guarnieri}

\affiliation{Department of Physics, University of Pavia, Via Bassi 6, 27100, Pavia, Italy}
\affiliation{Dahlem Center for Complex Quantum Systems,
Freie Universit\"{a}t Berlin, 14195 Berlin, Germany}

\author{Jens Eisert}
\affiliation{Dahlem Center for Complex Quantum Systems,
Freie Universit\"{a}t Berlin, 14195 Berlin, Germany}

\author{Harry J.~D. Miller}
\affiliation{Department of Physics and Astronomy, The University of Manchester, Manchester M13 9PL, UK}

\begin{abstract}
We consider a quantum system driven out of equilibrium via a small Hamiltonian perturbation. Building on the paradigmatic framework of linear response theory (LRT), we derive an expression for the full generating function of the dissipated work. Remarkably, we find that all information about the distribution can be encoded in a single  quantity, the standard relaxation function in LRT, thus opening up new ways to use phenomenological models to study non-equilibrium fluctuations in complex quantum systems. Our results establish a number of refined quantum thermodynamic constraints on the work statistics that apply to regimes of perturbative but arbitrarily fast protocols, and do not rely on assumptions such as slow driving or weak coupling. Finally, our approach uncovers a distinctly quantum signature in the work statistics that originates from underlying zero-point energy fluctuations. This causes an increased dispersion of the probability distribution at short driving times, a feature that can be probed in efforts to witness non-classical effects in quantum thermodynamics.
\end{abstract}

\maketitle

At the microscopic level, the traditional laws of thermodynamics fall short at providing an accurate description due to the fact that fluctuations in work, heat and entropy production play a preponderant role~\cite{Seifert2012,Lebowitz1999,Groot1961}.
Besides being of fundamental importance, these thermodynamic fluctuations have several direct implications on the performance of small-scale engines and (bio-)chemical reactions ~\cite{Seifert2015,Horo20a,Pietzonka2018,holubec2021fluctuations}. For these reasons, understanding the statistical aspects of dissipative processes has been the one of the overarching themes of the field of \emph{stochastic thermodynamics}~\cite{Onsager1931,Green1952,Kubo1966}. 
At even smaller scales, \emph{quantum mechanics} represents an additional source of fluctuations even in absence of any thermal agitation. While quantum properties have often been shown to lead to advantages over classical counterparts with regard to expectation values or speed-ups in total process time \cite{brandner2015c,jaramillo2016,campaioli2017}, many open questions still remain surrounding the thermodynamic cost associated to
quantum fluctuations (i.e., the process precision)~\cite{Guarnieri2019,Timpanaro2023,Hasegawa2020,Van_Vu2020b,Vo2022,Van_Vu2022,Van_Vu2023,Hasegawa2023,Miller2020,brenes2022particle,timpanaro2019,miller2021}.

One pressing question is to understand the nature of quantum fluctuations in \textit{finite-time} thermodynamic processes. Recent studies of slowly driven quantum systems have uncovered a wealth of strong results in this direction, ranging from finite-time thermodynamic bounds and trade-offs \cite{Cavina2017,Abiuso2020,bhandari2020,Brandner2020,eglinton2022,miller2021,alonso2022geometric}, general optimal control strategies \cite{Scandi,Miller2019}, geometric phase effects \cite{campisi2012geometric} and broad identifications of quantum signatures in work statistics \cite{Scandi2019,onishchenko2022probing}. However, going beyond slow driving regimes remains a significant challenge, due to the fact that finite time processes require precise knowledge about the underlying dynamics. This is especially difficult to obtain when a system is driven via time-dependent Hamiltonian driving or in contact with an external environment. One way around this is through \textit{linear response theory} (LRT), which allows one to utilise phenomenological models to make thermodynamic predictions based on the system's response to small perturbations. Since its original formulation by Kubo ~\cite{Kubo1957}, LRT has remained an indispensable tool for studying systems close to equilibrium \cite{marconi2008fluctuation}, with significant applications to quantum transport \cite{fu1993quantum}, many-body quantum physics \cite{suzuki1968dynamics} and quantum field theory \cite{calzetta2009nonequilibrium}.  In the context of quantum thermodynamics \cite{Topical}, one use of LRT has been the development of general optimisation strategies for minimal average work dissipation protocols \cite{Deffner2015c,Deffner2018,Deffner2018a}. 

In this paper, we develop a broader picture by characterizing the full quantum work distribution in LRT. In particular, we derive a universal and model-independent expression for all the statistical cumulants (average, variance, and higher fluctuations) of the dissipated work spent when driving a quantum systems out of equilibrium by means of a finite-time, weak perturbation. Crucially, we show that all these higher order fluctuations can be directly obtained through a single well-known quantity; the system's \emph{relaxation function}~\cite{Kubo1957,Deffner2018}. 
The latter is one of the central quantities in LRT and often represents the basis for phenomenological thermodynamic descriptions of complex many-body systems. Our main result opens new routes to analyze all the properties of work statistics in complex systems by means of this easily accessible quantity, as we  exemplify with examples of systems undergoing overdamped and underdamped Brownian motion. 

This result allows us to identify a new set of refined quantum thermodynamic constraints, including a fluctuation theorem and positivity of all work cumulants. These  solely rely on minimal assumptions such as unitary dynamics on the full system and  small Hamiltonian perturbations. Operationally speaking, our theory moreover predicts the existence of a distinctly quantum effect on the work probability distribution, which results in a significant broadening of the dispersion at low temperatures and prevents saturation of the thermodynamic uncertainty relation. We show that this \emph{non-classical signature} can be deeply connected to the breakdown of the equipartition theorem in quantum statistical mechanics. 

\textit{Quantum dissipated work statistics in LRT.} We begin by considering a quantum system unitarily driven out of equilibrium by means of a time-dependent Hamiltonian driving, $H_t:=H_0+\lambda_t V$, over a finite interval $t\in[0,\tau]$. Here, 
$t\mapsto \lambda_t$ is a dimensionless function characterising a particular driving protocol, and the operator $V$ is treated as a perturbation that is turned on at time $t=0$ (i.e.,~$\lambda_0=0$). 
The 
system is initially prepared in a thermal Gibbs state $\rho_0=\pi_0$ at inverse temperature $\beta=1/(k_B T)$, where we denote $\pi_t:=e^{-\beta H_t}/\mathcal{Z}_t$. 
After the driven evolution, the final state is given by $\rho_\tau=U_\tau \pi_0 U^\dagger_\tau$, with $U_\tau=\overleftarrow{\mathcal{T}}\text{exp}(i/\hbar\int^\tau_0 dt ' H_{t'})$. 
The main thermodynamic quantity of interest is the \emph{dissipated work} irreversibibly spent to drive the system out of equilibrium
\begin{align}\label{eq:dissipatedwork}
    \Wdiss:=W-\Delta F,
\end{align}
with $W$ denoting the stochastic quantum work defined through a two-time projective energy measurements at the beginning and end of the driving ~\cite{talkner2007fluctuation,esposito2009nonequilibrium}, and with 
$\Delta F=-\beta^{-1} \text{ln}(Z_\tau/Z_0)$ being the change in equilibrium free energy. 
A full stochastic thermodynamic description of the process can be derived from the resulting distribution in dissipated work $P(\Wdiss)$. As shown in Ref.~\cite{Guarnieri2018a}, this information is quantified by the \emph{quantum Renyi divergence} between the instantaneous equilibrium state $\pi_\tau$ and the non-equilibrium state $\rho_\tau$, since the 
\emph{cumulant generating function} (CGF) of the process is found to be
\begin{align}\label{eq:CGF}
    K(\eta):=\text{ln}\big\< e^{-\eta \beta\Wdiss}\big\>=(\eta-1)S_{\eta}(\pi_\tau||\rho_\tau),
\end{align}
where $S_\alpha(\rho_1||\rho_2):=(\alpha-1)^{-1}
\text{ln} \ \tr{\rho_1^\alpha \rho_2^{1-\alpha}}$ is the Renyi divergence of order $\alpha>0$, generalizing the quantum relative entropy. From the CGF, 
we can derive cumulants using the formula $\kappa^k_W :=(-k_B T)^k \lim_{\eta\to 0} {\partial^k  K(\eta)}/{\partial \eta^k}$.

In LRT one assumes a weak perturbation such that $|\lambda_t|\ll 1$ for $\forall t\in[0,\tau]$, with normalization $\|V\|=1$, resulting in a small deviation from the initial equilibrium state at all times given by \cite{bruus2004many}
\begin{align}\label{eq:kubo}
    \rho_t= \pi_0-\frac{i}{\hbar}\int^t_0 dt' \ \lambda_{t'}[V(t-t'),\pi_0].
\end{align}
with notation $A(t):=e^{iH_0 t/\hbar}A e^{-iH_0 t/ \hbar}$ indicating the interaction picture. 

Under this approximation, it is known that the linear-order correction to the average dissipated work is given in terms of the two-time integral ~\cite{andrieux2008,Deffner2015c,Deffner2018}
\begin{align}\label{eq:av}
\beta\langle \Wdiss \rangle=\frac{1}{2}\int^\tau_0 dt \int^\tau_0 dt'  \ \Psi_0(t-t')\dot{\lambda}_t \dot{\lambda}_{t'}.
\end{align}
Here $t\mapsto \Psi_0(t)$ denotes a central object in LRT known as the \textit{relaxation function} \cite{Kubo1957}, which can be expressed using the 
\emph{Kubo covariance} as follows
\begin{align}\label{eq:kubocov}
\Psi_0(t):= \beta\int^\beta_0 ds \ \langle V(-i\hbar s)V(t)\rangle_0-\beta^2 \langle V \rangle_0^2,
\end{align}
where $\langle .\rangle_0$ denotes the average with respect to the thermal state $\pi_0$. Physically, in LRT $\Psi_0$ allows to introduce a characteristic timescale $\tau_R:=\int^\infty_0 dt \  \Psi_0(t)/\Psi_0(0)$ over which two-time correlations in $t\mapsto V(t)$ decay in time. While the calculation of the relaxation function in principle requires knowledge of the exact dynamics of the system, the power of LRT lies in the fact that one may often use phenomenological models of $\Psi_0$  to investigate the generic behaviour of systems where this information may not be available. 
To do this one must impose certain constraints on any \textit{ansatz} that would ensure both dynamical and thermodynamic consistency. The two key properties we require are
\begin{align}\label{eq:constraint}
\Psi_0(t)=\Psi_0(-t), \, \forall t \ \ \  \ \tilde{\Psi}_0(\omega)\geq 0 \, \forall \omega,
\end{align}
where $\tilde{\Psi}_0(\omega)=\mathcal{F}[\Psi_0](\omega)$ denotes the Fourier transform of the relaxation function.
The first property reflects \emph{time-reversal symmetry} due to the underlying Hamiltonian dynamics. The second property in~\eqref{eq:constraint} expresses the positivity of its Fourier transform, and is equivalent to $\langle \Wdiss \rangle\geq 0$ in agreement with the second law of thermodynamics~\cite{Naze2020a}. Both conditions~\eqref{eq:constraint} follow from a fully Hamiltonian description and thus provide a consistency check for any approximate, phenomenological model of $\Psi_0(t)$. 

Going beyond average quantities is however necessary in order to properly characterize the thermodynamics of nano-scale processes, both classical and quantum even within LRT. In what follows we achieve this goal by systematically exploiting the state-expansion Eq.~\eqref{eq:kubo} in order to obtain the linear order corrections to the full cumulant 
generating function of the dissipated work Eq.~\eqref{eq:CGF}, and consequently to all its higher statistical cumulants $\kappa^k_W$.
Expanding the Renyi divergences in Eq.~\eqref{eq:CGF} up to second order in the perturbation strength, we arrive at our first main result for the full CGF in LRT and under time-reversal symmetry  $\Psi_0(t)=\Psi_0(-t)$,
\begin{align}\label{eq:CGFlinear1}
K(\eta):= -\int^\tau_0 dt \int^t_0 dt' \  \dot{\lambda}_t \dot{\lambda}_{t'}  \ \langle \langle \delta V(t),\delta V(t') \rangle \rangle_0^\eta    .
\end{align}
with $\delta A:=A-\tr{A \ \pi_0}$ (see Appendix A for details).
The bi-linear form in~\eqref{eq:CGFlinear1} denotes a generalised version of the Kubo correlation function Eq.~\eqref{eq:kubocov}
\begin{align}\label{eq:yCovariance}
\langle \langle A,B \rangle \rangle_0^\eta:=\int^{\beta\eta}_0 dx \ \int^{\beta-\beta x}_{\beta x} dy \ \langle  B(-i\hbar y) A\rangle_0,
\end{align}
Here the integration over imaginary time relates to the Green-Kubo-Mori-Zwanzig product \cite{fiorentino2023green}. This quantity originates from the field of quantum information-geometry \cite{petz1996monotone,petz2002covariance} and has recently found many successful applications in the field of stochastic thermodynamics, e.g. in slowly driven processes \cite{Scandi2019},
and in the context of the locality of temperature \cite{Locality}. 
It is however worth noting that, at variance with what happens in the slow driving regime where one neglects correlations over long times, the LRT accounts for memory effects due to finite-time driving, as reflected by the double time integral in Eq.~\eqref{eq:CGFlinear1}.

Our second main result establishes a link between Eq.~\eqref{eq:CGFlinear1} and the Kubo relaxation function $ \Psi_0$ Eq.~\eqref{eq:kubocov}. 
By using similar arguments of Ref.~\cite{Shitara2016}, one can re-express Eq.~\eqref{eq:CGFlinear1} as
\begin{align}\label{eq:CGFlinear2}
    K(\eta)= -\int^\tau_0 dt \int^\tau_0 dt' \, \dot{\lambda}_t \dot{\lambda}_{t'}\, [g_\eta \ast \Psi_0](t-t')
\end{align}
where
\begin{align}
g_\eta(t):=\mathcal{F}^{-1}\bigg[\frac{\text{Sinh}(\beta\hbar \omega(1-\eta)/2)\text{Sinh}(\beta\hbar\omega\eta/2)}{\beta\hbar\omega \ \text{Sinh}(\beta\hbar\omega/2)}\bigg](t),
\end{align}
is the inverse Fourier transform of a  model- and process-independent function.
The key insight of Eq.~\eqref{eq:CGFlinear2} is that the relaxation function now fully characterises the stochastic thermodynamics of a process for small perturbations, with $g_\eta$ acting as a universal generating function for the higher order fluctuations via its convolution with $\Psi_0$. 
The benefit of this expression is thus that one can now derive a number of general properties of the dissipated work statistics. Firstly, it is straightforward to see from Eq.~\eqref{eq:CGFlinear2} that the symmetry $K(\eta)=K(1-\eta)$ holds true for
all $\eta$. This in turn implies, via an inverse Laplace transform, the validity of the \emph{Evan-Searles fluctuation theorem} ~\cite{evans2002fluctuation} for $P(\Wdiss)$ 
\begin{align}\label{eq:FT}
    \frac{P(\Wdiss)}{P(-\Wdiss)}=e^{\beta \Wdiss}.
\end{align}
While this has been known to apply to systems driven slowly \cite{Scandi2019} or via time-symmetric driving protocols \cite{evans2008fluctuation}, here its validity is also demonstrated in LRT.  

In contrast to what is typically expected for linear response regimes in classical stochastic thermodynamics~\cite{jarzynski1997nonequilibrium,Speck,naze2023optimal}, our result predicts a quantum work distribution that is distinctly \textit{non-Gaussian} at finite temperatures. This can be seen by showing that the cumulants higher than the variance (i.e. skewness, kurtosis, etc) are non-zero. Taking derivatives of Eq.~\eqref{eq:CGFlinear2} and using the convolution theorem, we obtain that \emph{all} cumulants in dissipated work 
are positive and given by
\begin{align}\label{eq:cumulants}
\beta^k\kappa^k_W=\int_{\mathbb{R}}\frac{d\omega}{\sqrt{2\pi}}\tilde{\Psi}_0(\omega)\gamma^k(\omega)\bigg|\int^\tau_0 dt \ \dot{\lambda}_t e^{i\omega t}\bigg|^2\geq 0,
\end{align}
where 
\begin{align}
\begin{aligned}
\gamma^k(\omega):= 
\left\{
\begin{aligned}
& \frac{1}{2}(\beta\hbar\omega)^{k-1}\text{Coth}(\beta\hbar\omega/2) && \text{if } k \ \text{even,} \\
& \frac{1}{2}(\beta\hbar\omega)^{k-1} && \text{if } k \ \text{odd.}
\end{aligned}
\right.
\end{aligned} 
\end{align}
This positivity is a consequence of $\tilde{\Psi}_0(\omega)\geq 0$. 
These relations provide a refined set of constraints on the shape of $P(\Wdiss)$, demonstrating the presence of non-Gaussian tails for values $\Wdiss> 0$. These right tails have notable physical implications, since their presence and magnitude quantifies an increased likelihood of large dissipation accompanying the realizations of the given driving protocol. An analogous non-Gaussian signature has been shown to occur in slowly driven systems~\cite{Scandi2019,Miller2020}, and also many-body systems driven in finite time across a phase transition \cite{zawadzki2023non}. We emphasise the quantum origin of these tails; classical Gaussian behaviour can be recovered by either taking a high temperature limit, or assuming a commuting  perturbation $[V,H_0]=0$ (see Appendix A.3).

\begin{figure}[htbp!]
\begin{center}
\begin{tikzpicture} 
  \node (img1)  {\includegraphics[width=0.9\linewidth]{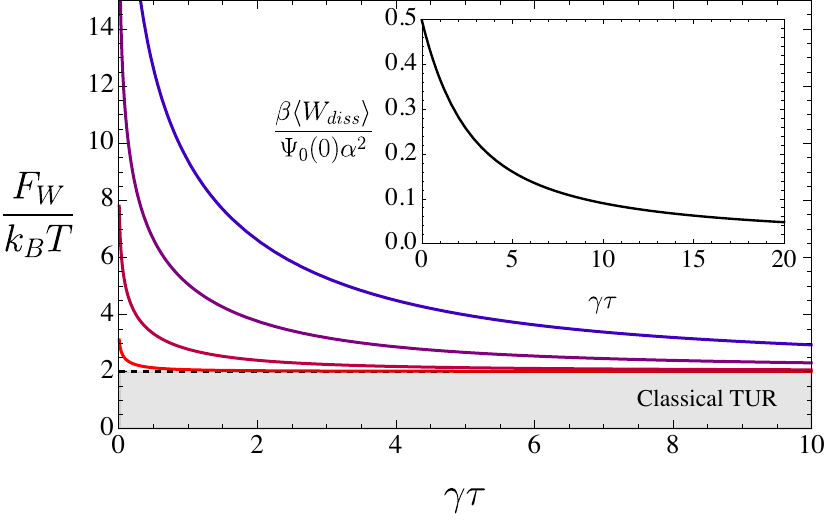}};
    \node[above=of img1, node distance=0cm, yshift=-1.7cm,xshift=-3.3cm] {{\color{black}{\bf{(a)}}}};
\end{tikzpicture}\\

\hspace*{-.2cm}
\begin{tikzpicture} 
  \node (img2)  {\includegraphics[width=0.89\linewidth]{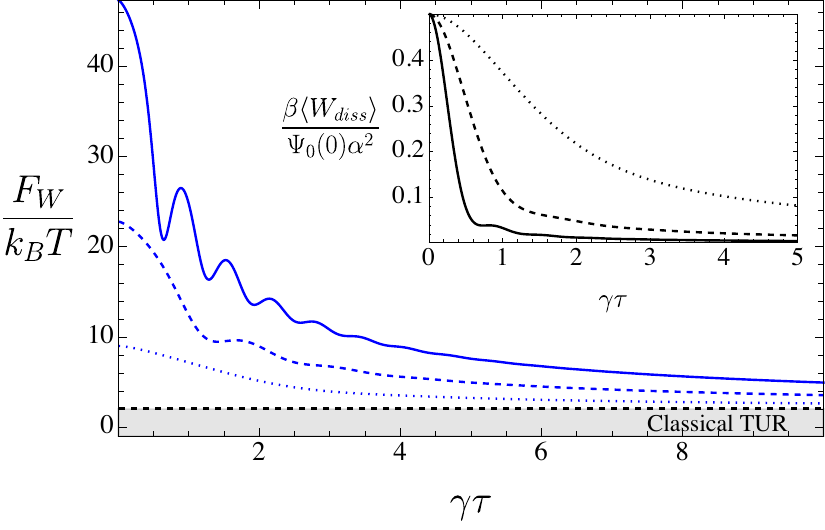}};
    \node[above=of img2, node distance=0cm, yshift=-1.7cm,xshift=-3.3cm] {{\color{black}{\bf{(b)}}}};
\end{tikzpicture}
\vspace*{-0.3cm}
\caption{
Fano factor~\eqref{eq:fano} for the work distribution 
in units of $k_B T$ as a function of the rescaled time $\gamma\tau$. Inset: Rescaled dissipated work $\beta\langle\Wdiss\rangle/(\Psi_0(0)\delta\lambda^2)$ as a function of the rescaled time $\gamma\tau$. Panel (a) depicts the overdamped model $\Psi_0^{(1)}$ in Eq.~\eqref{eq:models} and the other curves correspond to different values of $\hbar\beta\gamma = 0.5 \, (\text{blue}),2 \,  (\text{purple}), 5 \, (\text{magenta}), 10  \, (\text{red})$. Panel (b) refers to the underdamped model $\Psi_0^{(2)}$ and the curves correspond to fixed low temperature $\beta=5$ and different values of the frequency $\nu = 2\gamma \, (\text{blue dotted})$,
$5\gamma\, (\text{blue dot-dashed})$, 
$10\gamma \,(\text{blue dashed})$,
$15\gamma \,(\text{blue solid})$.}
\vspace*{-0.5cm}
\label{fig:example}
\end{center}
\end{figure}

\textit{Statistical interpretation of quantum signatures.} As a final result, our linear response approach also highlights a quantum signature in the statistics via a \emph{thermodynamic uncertainty relation}  (TUR). It is known that distributions satisfying the fluctuations relation~\eqref{eq:FT} are constrained by the TUR, which in the LRT regime can be expressed in terms of a lower bound 
\begin{align}\label{eq:fano}
    F_W:=\frac{\text{Var}(W)}{\langle \Wdiss \rangle}\geq 2k_B T
\end{align}
on the Fano factor (or relative dispersion) of the work distribution \cite{timpanaro2019}.
Classically, for a system that remains close to equilibrium such as in LRT \cite{hermans1991simple,naze2023optimal} or slow driving \cite{Speck,mandal2016analysis} one should expect to saturate the TUR. However, we show here that for finite temperatures the TUR cannot be saturated due to the influence of quantum fluctuations, and we provide a clear-cut statistical interpretation of this effect. To achieve this goal, we use $\tilde{\Psi}_0(\omega)$ to define a normalised probability distribution over the continuum of frequencies $\omega\in[0,\infty)$ associated with the system dynamics, which we refer to as \textit{pseudo-modes}:
\begin{align}\label{eq:probw}
    \tilde{P}(\omega):=\frac{\tilde{\Psi}_0(\omega)\big|\int^\tau_0 dt \ \dot{\lambda}_t e^{i\omega t}\big|^2}{\int^\infty_0 d\omega \ \tilde{\Psi}_0(\omega)\big|\int^\tau_0 dt \ \dot{\lambda}_t e^{i\omega t}\big|^2},
\end{align}
where the positivity is guaranteed by thermodynamic consistency Eq.~\eqref{eq:constraint}. Crucially, it now becomes apparent that the average energies of these modes determine the dispersion of the work distribution. First one can notice that, since $\Psi_0(t)=\Psi_0(-t)$, then $\tilde{\Psi}_0(\omega)$ is necessarily an even function of $\omega$. It is then straightforward to show from Eq.~\eqref{eq:cumulants} that the Fano factor Eq.~\eqref{eq:fano} can be expressed as
\begin{align}\label{eq:FQ}
    F_W=\bigg\langle \hbar\omega \ \text{Coth}(\beta\hbar \omega/2) \bigg\rangle_{ \tilde{P}}=2\langle \e_\omega \rangle_{ \tilde{P}},
\end{align}
where $\langle\cdot\rangle_{ \tilde{P}}$ denotes an average with respect to the pseudo-modes distribution $\omega\mapsto \tilde{P}(\omega)$ in Eq.~\eqref{eq:probw}. We can now recognise the RHS as twice the average total energy $\e_\omega$ of a  \emph{quantum harmonic oscillator} at frequency $\omega$ and in thermal equilibrium. 
Interestingly, this is still true even
if the original system at hand is 
not a harmonic system. This provides a remarkable and deep statistical-mechanical connection between the \textit{physical} dissipated-work fluctuations and the \textit{effective} energy distribution of the pseudo-modes associated with the driving protocol. 
Some key properties of this quantity can then be inferred, such as the following inequality
\begin{align}\label{eq:FQineq}
    F_W\geq \hbar\langle \omega \rangle_{ \tilde{P}} \text{Coth}(\beta\hbar \langle\omega\rangle_{ \tilde{P}}/2).
\end{align}
This follows from Jensen's inequality for the convex function $x
\mapsto x \ \text{Coth}(\beta x/2)$, and is tighter than the TUR~\eqref{eq:fano}. The genuine quantum origin of this effect is clear since the term $ \hbar \langle \omega \rangle$ represents in fact the average \textit{zero-point energy} of the pseudo-modes with respect to $\tilde{P}(\omega)$, and is responsible for preventing any saturation of the TUR. In fact, at low temperatures and short times we may approximate $\hbar \omega \text{Coth}(\beta\hbar\omega/2)\simeq  \hbar |\omega|$ meaning that $F_W\approx\hbar \langle \omega \rangle$ for $\beta\hbar/\tau_R\gg 1$ and hence the distribution will exhibit non-zero dispersion at absolute zero unlike a classical system. We can relate these non-vanishing fluctuations to the \textit{breakdown of the equipartition} theorem in quantum statistical mechanics \cite{feynman2018statistical}. Traditionally, this consideration applies to quantum systems in equilibrium and expresses the fact that energy cannot be equally shared amongst all degrees of freedom due to its discrete nature, implying a frequency dependence on the average energy rather than the classical prediction of $\langle \e_\omega\rangle=kT$ \cite{bialas2019quantum}. In the present context, we see that a similar breakdown occurs for non-equilibrium processes in linear response, preventing the saturation of the TUR.  

Finally, consistency with classical thermodynamics is ensured in the high-temperature/long-time limit $\beta\hbar/\tau_R\ll 1$, since $x \ \text{Coth}(x)\simeq  1$ for $x\ll 1$ and one recovers $F_W\rightarrow 2k_B T$. 
While a similar-in-spirit quantum signature has been derived for slowly-driven \emph{Markovian systems}  ~\cite{Miller2019,Scandi2019}, the result obtained here go significantly beyond 
that as it applies to arbitrary processes driving a system out of equilibrium in finite-time regimes, provided that the perturbation to the local 
Hamiltonian remains weak.

\textit{Examples.} As mentioned above, one of the most powerful consequences of our main result Eq.~\eqref{eq:CGFlinear2} is that it allows to characterize the full statistics of dissipation in more complex systems, where we may not have solutions to the full Hamiltonian dynamics. This is especially relevant for driven open quantum systems, whose dissipated work's statistics can still be described using our Eq.~\eqref{eq:CGFlinear2} provided one is simply given a good ansatz or approximation for the relaxation function. 
Here we benchmark this by considering the following two phenomenological models of $\Psi_0$, which are thermodynamically consistent according to Eq.~\eqref{eq:constraint}, 
\begin{align}\label{eq:models}
    &\Psi^{(1)}_0(t):=\Psi^{(1)}_0(0)e^{-\gamma |t|}, \\ &\Psi^{(2)}_0(t):=\Psi^{(2)}_0(0)e^{-\gamma |t|}\big(\text{cos}(\nu t)+(\gamma/\nu)\text{sin}(\nu |t|)\big).
\end{align}
In practice, such models can be built from the Kramers-Kronig relations and sum rules by imposing a number of Hamiltonian constraints that adequately characterise the system behaviour~\cite{Kubo1972}. 
In the first instance, $\Psi^{(1)}_0(t)$ provides a reasonable description of an overdamped Brownian dynamics \cite{Deffner2018}, with free parameter $\gamma=1/\tau_R$ setting the characteristic timescale over which the relaxation function decays. It also can describe many-body quantum systems such as the two-dimensional Ising model in a transverse field \cite{hotta2023quantum}. The second model, i.e., $\Psi^{(2)}_0$, 
includes an oscillatory behaviour with an additional degree of freedom $\nu$ that quantifies the frequency of an external potential, and should be viewed as a model of underdamped Brownian motion \cite{Deffner2018}. This model arises, for example, from the quantum dynamics of weakly-interacting magnetic systems \cite{white1983quantum}.  A remarkable feature of the linear response regime is that we can predict how the dispersion of the dissipated work distribution changes over time independent of the system-specific features contained in $\Psi_0(0)$. In Fig.~\ref{fig:example} (a) we plot the dispersion of the overdamped model as a function of time for different temperatures. In this case, we see that the correction \textit{monotonically} decays in the long time limit $\gamma\tau\to\infty$, indicating that the quantum fluctuations become less relevant at long times. Conversely, at short times we see a dramatic quantum signature with large dispersion in the work distribution above the classical FDR $F_W=2k_B T$. While long time decay can be ensured via the  damping terms in~\eqref{eq:models}, monotonicity is not necessarily guaranteed. This is clearly seen in the underdamped dynamics in Fig.~\ref{fig:example} (b), which indeed shows non-monotonic changes in the dispersion at short times. In Appendix B, we also consider a non-exponential model of $\Psi_0(t)$, and use the bound~\eqref{eq:FQineq} to derive analytic predictions for the Fano factor.

\textit{Conclusions.} In summary, we have systematically extended a linear response analysis to characterize full distribution for the quantum dissipated work statistics done along weakly perturbed processes. Our general formula for the CGF Eq.~\eqref{eq:CGFlinear2} paves new ways for studying properties of higher order work statistics of complex systems via phenomenological models of the relaxation function. This can be used to explore thermodynamic optimisation problems that revolve around stochastic fluctuations such as Pareto optimal work extraction~\cite{solon2018phase} and free energy estimation~\cite{blaber2020skewed}. The precise connection between the work distribution and relaxation function could also be used to explore the impact of non-Markovianity~\cite{Strasberg2018a} and phase transitions~\cite{Fei2020} on work statistics from a general perspective. Our results have also predicted a clear quantum signature~\eqref{eq:FQ} that causes an increase in the dispersion of the work distribution at short driving times and finite temperatures, and is universally applicable to composite systems that are driven by local perturbations. Since the Fano factor $F_W$ can be measured experimentally from the work statistics \cite{cerisola2017using}, this offers a clear route to detecting truly non-classical behaviour from thermodynamic variables. Finally, it would be interesting to explore generalisations of our formalism to study more general non-equilibrium states \cite{dechant2020fluctuation}.

\emph{Acknowledgements:} G.~G.~acknowledges fundings from European Union's Horizon 2020 research and innovation programme under the Marie 
Sklodowska-Curie Grant Agreement INTREPID, No.~101026667. J.~E. 
thanks the DFG (FOR 2724) and the 
FQXI. H.M. acknowledges funding from a Royal Society Research Fellowship (URF/R1/231394), and from the Royal Commission for the Exhibition of 1851.

\bibliographystyle{apsrev4-1}
\bibliography{blah.bib}

\appendix

\widetext

\section{Expression for the cumulant generating function for the dissipated work in linear response regime}

\

In this section, we will detail the derivation of the two expression for the \emph{cumulant generating function} (CGF) of the dissipated work in linear response  regime of the main text, i.e., Eqs.\ (8)-(10)). We first start by fixing the notation and introduce some of the main quantities used below. While we keep $\hbar$ explicit into the main text, we will set $\hbar = 1$ in this section in order to ease the notation. Let us start by considering a quantum system initially prepared in a thermal state with respect to a given Hamiltonian $H_0$, i.e., 
\begin{equation}
    \rho_0 = \pi_0 := \frac{e^{-\beta H_0}}{Z_0},
\end{equation}
where $\beta>0$ denotes the inverse temperature of the prepared state. From $t=0$ onwards, the Hamiltonian $H_t=H_0+\lambda_t V$ is externally driven, up to some arbitrary time $\tau$, by means of a driving schedule 
\begin{equation}
t\mapsto \lambda_t = \alpha g_t 
\end{equation}
($\lambda_0 = 0$), where $\alpha$ denotes the intensity of the perturbation and $t\mapsto g_t$ describes a generic time-dependent profile. As normalization. we pick the operator norm of the perturbation to take unity value, $\|V\|=1$.
Before we make any linear response approximation, we need to express the CGF $K(\eta)=\ln \langle e^{-\eta \beta\Wdiss}\rangle$ in a time local form, for $\eta>0$. This can be found by taking the time derivative of Renyi divergence, which has been determined in Ref.~\cite{Scandi2019} (see 
Eqs.~(I1)-(I10)) to be
\begin{align}\label{eq:CGFScandi}
    K(\eta)&=(\eta-1)S_{\eta}(\pi_\tau||\rho_\tau) = \int_0^\tau dt \, \frac{d}{dt}\left(\mathrm{Tr}\left[\pi_t^{\eta} \rho_t^{1-\eta}\right]\right)\notag \\ 
\nonumber &= -\beta\int_0^\tau dt \, \int_0^\eta   \left[\frac{\mathrm{Tr}\left[ \pi_t^{x}(\Delta_t \dot H_t)  \rho_t^{1-x}\right]}{\mathrm{Tr}\left[  \pi_t^{\eta} \rho_t^{1-\eta}\right]} \; +\int_0^x  dy \,\frac{\mathrm{Tr}\left[ \pi_t^{y}(\Delta_t \dot H_t) \pi_t^{x-y}(\log \pi_t - \log \varrho_t ) \varrho_t^{1-x}\right]}{\mathrm{Tr}\left[ \pi_t^{\eta} \varrho_t^{1-\eta}\right]}\right] \\
 &=: K_1(\eta) + K_2(\eta),
\end{align}
where $\Delta_t \dot{H}_t:=\dot{H}_t-\mathrm{Tr}[\dot{H}_t\pi_t]$, and for later convenience we have separated the two terms in the sum.

In this work, we will not put any restriction on the form of the protocol $t\mapsto \lambda_t$ nor on the speed of the driving (such as, e.g., in Refs.~\cite{Scandi2019,miller2021}), but rather we will assume that the perturbation strength is small, i.e., $\alpha\ll 1$. This means that we will perform a Taylor expansion in $\alpha$ of all the quantities entering Eq.~\eqref{eq:CGFScandi} and truncate to first order. This formal
framework is known as \textit{linear response} theory (LRT). 

\subsection{Derivation of Eq.\ (7)}

The first key ingredient from LRT is the linear order approximation for the instantaneous state of the system, which follows from a Dyson expansion of the unitarily evolved state $\rho_t:=U_t \pi_0 U_t^\dagger$ \cite{bruus2004many},
to get, in this approximation,
\begin{align}
    \rho_t\simeq\pi_0+\delta \rho_t, \ \ \ \ \ \delta\rho_t=-i \ \alpha \int^t_0 dt' \  g_{t'} [V(t-t'),\pi_0],
\end{align}
where $V(t):=e^{i t H_0}V e^{-it H_0}$ denotes the perturbation evolved in the interaction picture. As well as this, we need to also consider how perturbation effects the instantaneous equilibrium state as well. The Duhamel identity provides a mean of determining this, which gives the Taylor expansion 
\begin{align}
    e^{A+\alpha B}=e^A+\alpha\int^1_0 dy \ e^{s A} B e^{(1-s)A}+\mathcal{O}(\alpha^2),
\end{align}
for an exponential operator \cite{Kubo1957}.
It then follows that the thermal state at time $t$ is given by
\begin{align}
    \pi_t:=\frac{e^{-\beta (H_0+\alpha g_t V)}}{\tr{e^{-\beta (H_0+\alpha g_t V)}}}=\pi_0-\alpha\beta g_t \mathbb{J}_0(\delta V) + \mathcal{O}(\alpha^2),
\end{align}
where we have introduced the superoperator 
\begin{equation}\label{eq:J}
    \mathbb{J}_0(.):=\int^1_0 dy \ \pi_0^y \  (.)\pi_0^{1-y},
\end{equation}
and have defined the shifted operator $\delta V:=V-\tr{V\pi_0}$. From this point on, in order to not over-cumber the notation, all the $'='$ signs in equations (unless explicitly stated otherwise) will signify an equality up to terms of the order $\mathcal{O}(\alpha^2)$. 
Analogous calculations allow to obtain analytic expressions for the other quantities of interest entering Eq.~\eqref{eq:CGFScandi}, such as 
the following.
\begin{itemize}
\item The matrix logarithm of the instantaneous state \cite{hiai2014introduction},
\begin{equation}
    \log \rho_t \simeq \text{log} \ \pi_0+\mathbb{J}_0^{-1}(\delta \rho),
\end{equation}
with $\mathbb{J}^{-1}_0$ being the inverse superoperator of $\mathbb{J}_0$, i.e., $\mathbb{J}_0^{-1} \mathbb{J}_0(.) = \mathbb{J}_0\mathbb{J}_0^{-1}(.) = (.)$ .
\item Their matrix powers \cite{Scandi2019}
\begin{align}
    \rho_t^x = (\pi_0+\delta \rho)^x = \pi_0^x+\mathbb{J}_0^{(x)} \left(\mathbb{J}_0^{-1}(\delta\rho)\right), \, \, \, \,   \pi_t^x = \pi_0^x-\beta\lambda_t\mathbb{J}_0^{(x)}(\delta V),
\end{align}
where we have introduced the superoperator 
\begin{equation}
    \mathbb{J}_0^{(x)}(.):=\int^x_0 dy \  \pi_0^y \  (.)\pi^{x -y}_0,
\end{equation}
consistent with Eq.~\eqref{eq:J}, i.e., $\mathbb{J}_0(.) \equiv \mathbb{J}^{(1)}_0(.)$;
\item The instantaneous deviation of the Hamiltonian from its thermal expectation value 
\begin{align}
    \Delta_t \dot{H}_t =  \dot{\lambda}_t \  \delta V + \beta \dot{\lambda}_t \lambda_t \ \text{cov}^1(V,V),
\end{align}
where we have introduced the $y$-covariance between two operators $A$ and $B$ with respect to $\pi_0$ by
\begin{align}\label{eq:yCovariance}
\text{cov}^y\big(A,B\big):=\tr{\pi_0^y \ \delta A \ \pi_0^{1-y} \ \delta B},
\end{align}
with $\delta A=A-\tr{A \pi_0}$.
\end{itemize}
Making use of these expressions, it is straightforward to show that the denominator of Eq.~\eqref{eq:CGFScandi} reduces in linear response regime to
\begin{equation}
    \mathrm{Tr}\left[  \pi_t^{\eta} \rho_t^{1-\eta}\right] = 1,
\end{equation}
where we remind the reader that we implicitly neglected the terms $\mathcal{O}(\alpha^2)$.
Further calculations allow to express the two terms in the expression for the CGF of the dissipated work Eq.~\eqref{eq:CGFScandi} as
\begin{align}
    &K_1(\eta)=-\frac{\beta^2 \eta \lambda_\tau^2}{2} \ \int^1_0 dy \ \text{cov}^y(V,V)+\beta^2\lambda^2_\tau \int^\eta_0 dx \int^{x}_0 dy \ \text{cov}^y\big(V,V\big), \label{eq:K1}\\
    &K_2(\eta)=-\beta\int^\tau_0 dt \ \dot{\lambda}_t \int^\eta_0 dx \int^{1-x}_x dy \  \text{cov}^y\big(\mathbb{J}_0^{-1}(\delta\rho),V\big).
\end{align}
A further insightful simplification stems from expanding the operators appearing in $K_2(\eta)$ onto the eigenbasis of $H_0=\sum_n \epsilon_n\ket{n}\bra{n}$: Setting $M=\mathbb{J}_0^{-1}(\delta V)$ for convenience, one has
\begin{align}\label{eq:K2}
    \nonumber K_2(\eta)&=-\beta \int^\eta_0 dx \int^{1-x}_x dy \int^\tau_0 dt \ \dot{\lambda}_t \int^t_0 dt' \ \lambda_{t'} \  \phi_y(t-t') \\
    &=-\beta \int^\eta_0 dx \int^{1-x}_x dy \int^\tau_0 dt \ \dot{\lambda}_t \int^t_0 dt' \ \lambda_{t-t'} \  \phi_y(t'),
\end{align}
where we have defined
\begin{align}
    \phi_y(v):=i\sum_{n,m}   \ p_n^y \ p_m^{1-y}e^{-iv \beta^{-1}\text{log}(p_n/p_m)}(p_n-p_m) V_{n,m} M_{m,n},
\end{align}
and have diagonalised the thermal state $\pi_0=\sum_n p_n \ket{n}\bra{n}$ with $p_n\propto e^{-\beta E_n}$ and set $\Delta E_{n,m}:=\epsilon_n-\epsilon_m$. If we now consider its integral
\begin{align}
    \nonumber\psi_y(t')=\int^{t'}_0 dt'' \ \phi_y(t'')&=-\beta \sum_{n,m}   \ p_n^y \ p_m^{1-y}\bigg(1-e^{it'\Delta E_{n,m}}\bigg)\frac{(p_n-p_m)}{\text{log}(p_n/p_m)} V_{n,m} M_{m,n}, \\
    \nonumber&=\beta \sum_{n,m}   \ p_n^y \ p_m^{1-y}\bigg(\bra{n}\mathbb{J}_0[V]\ket{m}-\bra{n}\mathbb{J}_0[V(t')]\ket{m}\bigg)\bra{m}\mathbb{J}_0^{-1}[\delta V]\ket{n}, \\
    \nonumber&=\beta\tr{\pi^y V\pi^{1-y}\delta V}-\beta\tr{\pi^y V(t')\pi^{1-y}\delta V}, \\
    &=\beta\text{cov}^y\big(V,V\big)-\beta\text{cov}^y\big(V(t'),V\big),
\end{align}
we can use this expression in Eq.~\eqref{eq:K2} once an integration by parts is performed, leading to
\begin{align}
    \nonumber K_2(\eta)&=-\beta \int^\eta_0 dx \int^{1-x}_x dy \int^\tau_0 dt \ \dot{\lambda}_t \bigg(\lambda_0 \psi_y(t)-\lambda_t \psi_y(0)-\int^t_0 dt' \ \frac{d\lambda_{t-t'}}{dt'} \  \psi_y(t')\bigg) \\
    \nonumber &=\beta \int^\eta_0 dx \int^{1-x}_x dy \int^\tau_0 dt \ \dot{\lambda}_t \int^t_0 dt' \ \frac{d\lambda_{t-t'}}{dt'} \  \psi_y(t'), \\
    &=-\beta^2\int^\tau_0 dt \ \dot{\lambda}_t \int^t_0 dt \  \dot{\lambda}_{t-t'} \int^\eta_0 dx \int^{1-x}_x dy \  \text{cov}^y\big(V(t'),V\big)+\frac{\beta^2\lambda_\tau^2}{2}\int^\eta_0 dx \int^{1-x}_x dy \  \text{cov}^y\big(V,V\big).
\end{align}
To get the second line of the equation above, we have used the fact that $\lambda_0 = 0$ and $\psi_y(0) = 0$.
Finally, we make use of this expression for $K_2(\eta)$ and combine it with $K_1(\eta)$ from Eq.~\eqref{eq:K1}, and then exploit the fact that
\begin{align}
    \bigg(\int^x_0-\frac{1}{2}\int^1_0+\frac{1}{2}\int^{1-x}_x\bigg) dy \  \text{cov}^y\big(V,V\big)=0
\end{align}
due to the symmetry $\text{cov}^y\big(V,V\big)=\text{cov}^{1-y}\big(V,V\big)$. This lead us to obtain the following compact form for the CGF of the dissipated work in linear response regime
\begin{align}\label{eq:CGFapp1}
    K(\eta)&=-\beta^2\int^\tau_0 dt \ \dot{\lambda}_t\int^t_0 dt' \ \dot{\lambda}_{t'} \int^\eta_0 dx \int^{1-x}_x dy \ \text{cov}^y \big(V(t-t'),V(0)\big) \notag\\
    \nonumber &= -\beta^2\int^\tau_0 dt\int^t_0 dt' \ \dot{\lambda}_t\dot{\lambda}_{t'} \int^\eta_0 dx \int^{1-x}_x dy \ \text{cov}^y \big(V(t),V(t')\big), \\
    &= -\beta^2\int^\tau_0 dt \int^t_0 dt' \  \dot{\lambda}_t \dot{\lambda}_{t'}  \ \langle \langle V(t),V(t') \rangle \rangle_0^\eta 
\end{align}
where we have used the time translational invariance of the covariance appearing in the integrand. This represents the first expression of the main result Eq. (8).

\subsection{Derivation of Eq.\ (9)}

\noindent In this section, we prove that our formula for the CGF~\eqref{eq:CGFapp1} can be rewritten in terms of the \textit{relaxation function}, defined by
\begin{align}\label{eq:relaxapp}
    \nonumber \Psi_0(t-t')&:= \beta\int^\beta_0 ds \ \langle V(-i s)V(t-t')\rangle_0-\beta^2 \langle V \rangle_0^2 \\
    &=\beta^2 \int^1_0 dy \ \text{cov}^y\big(V(t),V(t')\big).
\end{align}
Our approach will be based on a method used by Shitara and Ueda  who found an elegant connection between the linear response function and families of quantum covariances \cite{Shitara2016}.   Let us first define the Fourier transform 
\begin{equation}\label{eq:FourierCov}
    \mathcal{C}_{\eta}(\omega) := \frac{1}{\sqrt{2\pi}}\int_{\mathbb{R}}d (t-t')\, e^{i\omega (t-t')} \langle\langle V(t),V(t') \rangle\rangle^\eta_{0}
\end{equation}
of the bilinear form in Eq.~\eqref{eq:CGFapp1} with respect to time difference $t-t'$.
We also define a function $f_\eta$ 
for a parameter $\eta>0$ as
\begin{align}\label{eq:feta}
    f_\eta(x):=\frac{(x^\eta-1)(x^{1-\eta}-1)}{\text{log}^2(x)}.
\end{align}
The expansion of Eq.~\eqref{eq:FourierCov} on to the basis of the state $\pi_0$ gives
\begin{align}\label{eq:res1}
    \mathcal{C}_{\eta}(\omega) &= \frac{1}{\sqrt{2\pi}}\int_{\mathbb{R}}d(t-t') \,e^{i\omega (t-t')}\,  \sum_{n,m} \left(-\int^\eta_0 dx \int^{1-x}_x dy \, p_n^y \ \delta V_{n,m} e^{i(\epsilon_n-\epsilon_m)(t-t')}p_m^{1-y}\delta V_{m,n}\right)\notag\\&=  \frac{1}{\sqrt{2\pi}}\int_{\mathbb{R}}d(t-t') \ e^{i\omega (t-t')}\, \sum_{n,m} e^{i(\epsilon_n-\epsilon_m)(t-t')} \left[\int^\eta_0 dx \int^{x}_{1-x} dy \left(\frac{p_n}{p_m}\right)^y \right]\, \delta V_{n,m} p_m \delta V_{m,n}\notag\\
    &= \frac{1}{\sqrt{2\pi}}\sum_{n,m} \left( \int_{\mathbb{R}}d(t-t')\ e^{i(\omega-\beta^{-1}\text{log} (p_n/p_m))(t-t')} \right)\,\left[\frac{\left((p_n/p_m)^{\eta}-1\right)\left((p_n/p_m)^{1-\eta}-1\right)}{\text{log}^2(p_n/p_m)} \right] \delta V_{n,m} \  p_m \  \delta V_{m,n} \notag\\
    &=   \sum_{n,m} \sqrt{2\pi}\delta(\omega-\beta^{-1}\text{log}  (p_n/p_m))\,f_{\eta}(p_n/p_m) \delta V_{n,m} p_m \delta V_{m,n} \notag\\
    &= \sqrt{2\pi} f_{\eta}(e^{-\beta\omega})\sum_{n,m}\delta(\omega-\beta^{-1}\text{log}(p_n/p_m)) \delta V_{n,m} p_m \delta V_{m,n}, 
\end{align}
where we have used the relation $\epsilon_n-\epsilon_m = -\beta^{-1}\text{log} (p_n/p_m) $ and the definition of the Dirac delta $\int_{\mathbb{R}}dt\,e^{i\omega t} = 2\pi\delta(\omega)$ and its properties. The last term in the above expression, i.e., $\sum_{n,m}\delta(\omega-\beta^{-1}\text{log} (p_n/p_m)) \delta V_{n,m} p_m \delta V_{m,n}$, could be further simplified, but it will be more convenient to keep it in this form for the following calculations.
Let us now consider the Fourier transform of the relaxation function Eq.~\eqref{eq:relaxapp}, i.e., 
\begin{equation}\label{eq:FourierPsi0}
    \Psi_0(\omega) = \frac{\beta^2}{\sqrt{2\pi}}\int_{\mathbb{R}}d(t-t') \,e^{i\omega (t-t')}\, \Psi_0(t-t') =  \frac{\beta^2}{\sqrt{2\pi}}\int_{\mathbb{R}}d(t-t') \,e^{i\omega (t-t')}\, \text{cov} \left(V(t),V(t')\right).
\end{equation}
Analogous calculations can be made for this quantity
\begin{align}\label{eq:res2}
    (k_B T)^2\Psi_0(\omega) &= \frac{1}{\sqrt{2\pi}}\int_{\mathbb{R}}d(t-t') \,e^{i\omega (t-t')}\,  \sum_{n,m} \left(\int^{1}_0 dy \, p_n^y \ \delta V_{n,m} e^{i(\epsilon_n-\epsilon_m)(t-t')}p_m^{1-y}\delta V_{m,n}\right)\notag\\
    &= \frac{1}{\sqrt{2\pi}}\int_{\mathbb{R}}d(t-t') \, e^{i\omega (t-t')}\, \sum_{n,m} e^{i(\epsilon_n-\epsilon_m)(t-t')} \left[\int^{1}_0 dy \left(\frac{p_n}{p_m}\right)^y \right]\, \delta V_{n,m} p_m \delta V_{m,n}\notag\\
    &= \frac{1}{\sqrt{2\pi}}\sum_{n,m} \left(\int_{\mathbb{R}}d(t-t') \ e^{i(\omega-\beta^{-1}\text{log} (p_n/p_m))(t-t')} \right)\,\left[\frac{(p_n/p_m)-1}{\text{log}(p_n/p_m)} \right] \delta V_{n,m} p_m \delta V_{m,n} \notag\\
    &= \sqrt{2\pi} \ \frac{1-e^{-\beta\omega}}{\beta\omega}\sum_{n,m}\delta(\omega-\beta^{-1}\text{log} (p_n/p_m)) \delta V_{n,m} p_m \delta V_{m,n}.
\end{align}
By direct comparison between Eqs.~\eqref{eq:res1} and~\eqref{eq:res2} (and putting $\hbar$ back in),  we obtain the following result
\begin{equation}\label{eq:fetapower}
    \beta^2\mathcal{C}_{\eta}(\omega) = \frac{\beta\hbar\omega f_{\eta}(e^{-\beta\hbar\omega})}{(1-e^{-\beta\hbar\omega})}\Psi_0(\omega).
\end{equation}
Applying the inverse Fourier transform to both sides and using the convolution theorem gives us
\begin{align}
    \beta^2\langle\langle V(t),V(t') \rangle\rangle^\eta_{0}=\mathcal{F}^{-1}\bigg[\frac{\beta\hbar\omega f_{\eta}(e^{-\beta\hbar\omega})}{(1-e^{-\beta\hbar\omega})}\bigg]\ast\Psi_0(t-t').
\end{align}
Plugging this into~\eqref{eq:CGFapp1} and using the symmetry under time-reversal, $\Psi_0(t)=\Psi_0(-t)$, we get a new form for the CGF 
\begin{align}\label{eq:CGFfinal}
    \nonumber K(\eta)&=\int^\tau_0 dt \int^t_0 dt' \, \dot{\lambda}_t \dot{\lambda}_{t'}\, \mathcal{F}^{-1}\bigg[\frac{\beta\hbar\omega f_{\eta}(e^{-\beta\hbar\omega})}{(e^{-\beta\hbar\omega}-1)}\bigg]\ast\Psi_0(t-t') \\
    &=\frac{1}{2}\int^\tau_0 dt \int^\tau_0 dt' \, \dot{\lambda}_t \dot{\lambda}_{t'}\, \mathcal{F}^{-1}\bigg[\frac{\beta\hbar\omega f_{\eta}(e^{-\beta\hbar\omega})}{(e^{-\beta\hbar\omega}-1)}\bigg]\ast\Psi_0(t-t')
\end{align}
in terms of the relaxation function. 
Using the definition~\eqref{eq:feta}, a straightforward rearrangement confirms our main result Eq.\ (10).

\subsection{Recovering the classical Gaussian limit}

\

Here we will show two ways in which one may recover the classical statistics of work, which predicts a Gaussian work distribution of the form \cite{Jarzynski1997,naze2023optimal}
\begin{align}\label{eq:gaussian}
    P(\Wdiss)=\sqrt{\frac{\beta}{4\pi \langle \Wdiss \rangle}}e^{-\beta \frac{(\Wdiss-\langle \Wdiss \rangle)^2}{4\langle \Wdiss \rangle}}.
\end{align}
As stated in the main text, the cumulants in dissipated work obtained from the derivatives of the CGF~\eqref{eq:CGFfinal} take the form
\begin{align}\label{eq:cumulants}
\beta^k\kappa^k_W=\int_{\mathbb{R}}\frac{d\omega}{\sqrt{2\pi}}\tilde{\Psi}_0(\omega)\gamma^k(\omega)\bigg|\int^\tau_0 dt \ \dot{\lambda}_t e^{i\omega t}\bigg|^2,
\end{align}
where 
\begin{align}
\begin{aligned}
\gamma^k(\omega):= 
\left\{
\begin{aligned}
& \frac{1}{2}(\beta\hbar\omega)^{k-1}\text{Coth}(\beta\hbar\omega/2) && \text{if } k \ \text{even,} \\
& \frac{1}{2}(\beta\hbar\omega)^{k-1} && \text{if } k \ \text{odd.}
\end{aligned}
\right.
\end{aligned} 
\end{align}
The first notion of classicality occurs in the high temperature limit which can be found by expanding the cumulants for small $\beta\hbar$. Using $x \  \text{Coth} \ x\simeq 1+\mathcal{O}(x^2)$ and dropping terms of order $\mathcal{O}(\beta^2\hbar^2)$ we find $\gamma^1(\omega)\simeq \frac{1}{2}\gamma^2(\omega)$ and $\gamma^k(\omega)\simeq 0$ for $k\geq 3$. Formally this Taylor expansion is valid whenever  the temperature is small relative to the characteristic timescale of $\Psi_0(t)$, ie. $\beta\hbar \ll\tau_R$, since this governs the decay the larger modes $\omega$ contributing to $\tilde{\Psi}_0(\omega)$. From this approximation we can conclude that the work distribution is Gaussian with $\langle \Wdiss \rangle=\frac{1}{2}\beta\text{Var}(\Wdiss)$, in agreement with the form~\eqref{eq:gaussian}.

A second way to recover classical statistics is if the perturbation commutes with the bare Hamiltonian, $[V,H_0]=0$. In that case the relaxation function becomes time independent, $\Psi_0(t)=\Psi_0(0)$. Then the Fourier transform becomes singular, $\tilde{\Psi}_0(\omega)=\sqrt{2\pi}\delta(\omega)\Psi_0(0)$, and plugging this into~\eqref{eq:cumulants} we find
\begin{align}
     \beta\langle \Wdiss \rangle=\frac{1}{2}\beta^2 \text{Var}(\Wdiss)=\frac{1}{2}\lambda_\tau^2 \Psi_0(0), \ \ \ \text{and} \ \ \  \kappa^k_W=0 \ \forall k\geq 3
\end{align}
The vanishing skewness and proportionality between the first and second cumulants again recovers the Gaussian distribution~\eqref{eq:gaussian}.

\section{Examples of relaxation functions and their work dispersion  }

\

In this section, we look in greater detail a selection of different phenomenological models for the relaxation function and the corresponding dispersion of the work distribution.

\subsection{Overdamped versus underdamped models}

\

We first consider both the two models of $\Psi_0(t)$ presented in the main text, 
\begin{align}
    &\Psi^{(1)}_0(t)=\Psi^{(1)}_0(0)e^{-\gamma |t|} ,\\
    &\Psi^{(2)}_0(t)=\Psi^{(2)}_0(0)e^{-\gamma |t|}\big(\text{cos}(\nu t)+(\gamma/\nu)\text{sin}(\nu |t|)\big).
\end{align}
These models represent, respectively, overdamped and underdamped Brownian motion. Their Fourier transforms are
found to be
\begin{align}
    &\tilde{\Psi}^{(1)}_0(\omega)=\Psi^{(1)}_0(0)\sqrt{\frac{2}{\pi}}\frac{\gamma}{\gamma^2+\omega^2}, \\
    &\tilde{\Psi}^{(2)}_0(\omega)=\Psi^{(2)}_0(0)\sqrt{\frac{2}{\pi}} \ \frac{2\gamma (\nu^2+\gamma^2)}{\nu^4+2\nu^2(\gamma-\omega)(\gamma+\omega)+(\gamma^2+\omega^2)^2},
\end{align}
which are clearly positive as required by Eq. (7) in the main text, implying that we can use them as thermodynamically consistent linear response models. We choose a linear protocol
\begin{align}
    \lambda_t=\alpha \ t/\tau
\end{align}
so that
\begin{align}
    \bigg|\int^\tau_0 dt \ e^{i\omega t}\dot{\lambda}_t\bigg|^2=\frac{\alpha^2}{\tau^2} \bigg|\int^\tau_0 dt \ e^{i\omega t}\bigg|^2=\frac{2\alpha^2}{\tau^2\omega^2} \big(1-\text{Cos}(\tau\omega)\big)=\frac{4\alpha^2}{\tau^2\omega^2}\text{Sin}^2(\tau\omega/2) .
\end{align}
For $\omega\mapsto \tilde{\Psi}^{(1)}_0(\omega)$, we compute the average as
\begin{align}
    \beta\langle \Wdiss \rangle=\frac{\gamma\alpha^2}{2\pi\tau^2}\Psi^{(1)}_0(0)\int_{\mathbb{R}}d\omega \frac{4 \ \text{Sin}^2(\tau\omega/2)}{\gamma^2\omega^2+\omega^4}= \frac{\alpha^2 \Psi^{(1)}_0(0)}{ \gamma^2 \tau^2} \big(\gamma \tau +e^{-\gamma \tau}-1\big).
\end{align}
Going further, the variance is given by
\begin{align}
    \nonumber \beta^2\text{Var}(W)&=\frac{\gamma\alpha^2}{\pi\tau^2}\Psi^{(1)}_0(0)\int_{\mathbb{R}}d\omega \frac{4 \ \text{Sin}^2(\tau\omega/2)}{\gamma^2 \omega^2+\omega^4}(\beta \hbar\omega/2)\text{Coth}(\beta\hbar\omega/2), \\
    &=\frac{4\beta \hbar\gamma \ \alpha^2}{\pi\gamma^2\tau^2}\Psi^{(1)}_0(0)\int_{0}^\infty d\tilde{\omega} \frac{ \ \text{Sin}^2(\tau\gamma\tilde{\omega}/2)}{\tilde{\omega}+\tilde{\omega}^3} \text{Coth}(\beta\hbar\gamma\tilde{\omega}/2),
\end{align}
This integral cannot be done analytically for arbitrary $\beta>0$, so we proceed numerically. The key point to note is that the Fano factor will become independent of $\Psi_0(0)$ and depend only on the relaxation timescale $\gamma>0$ and temperature, with
\begin{align}
    \beta F_W(x,y)=\frac{8 x}{\big(y +e^{-y}-1\big)} \times \int_{0}^\infty d\tilde{\omega} \frac{ \ \text{Sin}^2(y \ \tilde{\omega}/2)}{\tilde{\omega}+\tilde{\omega}^3} \text{Coth}(x \ \tilde{\omega}/2)
\end{align}
 now being a function of dimensionless variables $x:=\hbar\beta\gamma$ and $y:=\tau \gamma$. For our second example, the average and variance become
\begin{align}
    &\beta\langle \Wdiss\rangle=\frac{8\alpha^2}{\pi\gamma^2\tau^2}\Psi^{(2)}_0(0)\int^\infty_0 d\tilde{\omega} \ \frac{(\tilde{\nu}^2+1)\text{Sin}^2(\tau\gamma\tilde{\omega}/2)}{\tilde{\omega}^2 \tilde{\nu}^4+2\tilde{\nu}^2(\tilde{\omega}-\tilde{\omega}^2)(\tilde{\omega}+\tilde{\omega}^2)+(\tilde{\omega}+\tilde{\omega}^3)^2)} ,\\
    &\beta^2\text{Var}(W)=\frac{8\alpha^2 \beta\hbar\gamma}{\pi\gamma^2\tau^2}\Psi^{(2)}_0(0)\int^\infty_0 d\tilde{\omega} \ \frac{(\tilde{\nu}^2+1)\tilde{\omega} \ \text{Coth}(\beta\hbar\gamma\tilde{\omega}/2)\text{Sin}^2(\tau\gamma\tilde{\omega}/2)}{\tilde{\omega}^2 \tilde{\nu}^4+2\tilde{\nu}^2(\tilde{\omega}-\tilde{\omega}^2)(\tilde{\omega}+\tilde{\omega}^2)+(\tilde{\omega}+\tilde{\omega}^3)^2)}
\end{align}
where we have set $\nu:=\gamma\tilde{\nu}$. These integrals must again be computed numerically. 

\subsection{Model with non-exponential relaxation}

\begin{figure}[htbp!] \includegraphics[width=0.89\linewidth]{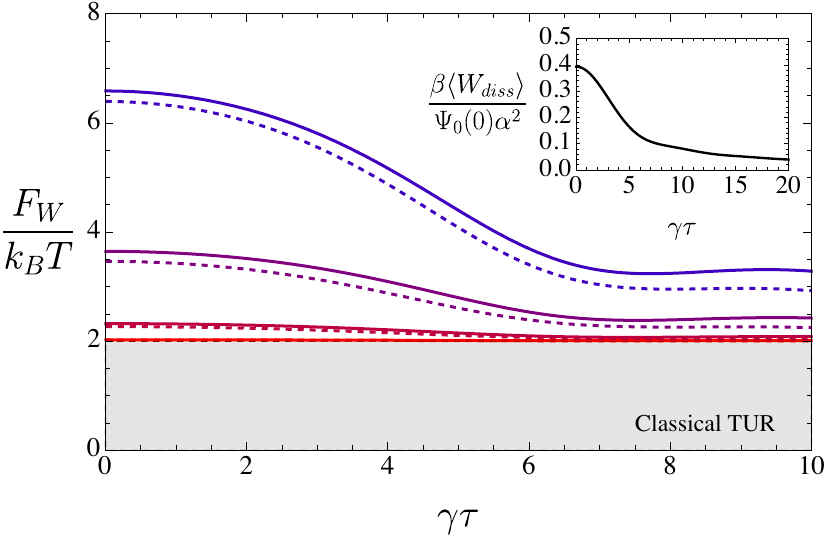};
\vspace*{-0.3cm}
\caption{
Fano factor (solid colored lines) for the model with non-exponential relaxation
in units of $k_B T$ as a function of the rescaled time $\gamma\tau$ and for different values of temperature: $\hbar\beta\gamma = 0.5 \, (\text{blue}),2 \,  (\text{purple}), 5 \, (\text{magenta}), 10  \, (\text{red})$. The dashed lines correspond to the respective analytic lower bounds Eq.~\eqref{eq:boundszeropointenergy}.Inset: Rescaled dissipated work $\beta\langle\Wdiss\rangle/\Psi_0(0)\delta\lambda^2$ as a function of the rescaled time $\gamma\tau$. }
\vspace*{-0.5cm}
\label{fig:newexample}
\end{figure}

\noindent An example of a non-exponential relaxation function that can arise, for example, in the Ising model \cite{glauber1963time,Naze2020a} is given by
\begin{align}
    \Psi_0^{(3)}(t)\propto J_0(\gamma t),
\end{align}
with $J_\nu(x)$ the Bessel function of the first kind. In Fourier space this is
\begin{align}
    \tilde{\Psi}_0(\omega)\propto\frac{ \left(1- \theta\left[\omega^2-\gamma^2 \right]   \right)}{\sqrt{\gamma^2 - \omega^2}}.
\end{align}
In this model, the average dissipation for a linear protocol becomes
\begin{align}
    \nonumber\beta\langle \Wdiss \rangle&\propto\frac{1}{\tau^2}\int_{\mathbb{R}}d\omega \frac{ \ \text{Sin}^2(\tau\omega/2)}{\sqrt{\gamma^2\omega^4 - \omega^6}}\left(1- \theta\left[\omega^2-\gamma^2 \right]   \right), \\
    \nonumber&\propto\frac{1}{\gamma^2\tau^2}\int^1_{0}d\omega \frac{ \ \text{Sin}^2(\gamma\tau\tilde{\omega}/2)}{\sqrt{\tilde{\omega}^4 - \tilde{\omega}^6}}, \\
    &=\frac{\pi}{8 \gamma\tau} \left( J_1(\gamma\tau) (-2 + \pi \gamma\tau H_0(\gamma\tau)) + \gamma\tau J_0(\gamma\tau) (2 - \pi H_1(\gamma\tau)) \right),
\end{align}
with $x\mapsto H_\nu(x)$ 
being the Struve function. The Fano factor is then
\begin{align}
    \nonumber F_W&=[\frac{\pi}{8 \gamma\tau} \left( J_1(\gamma\tau) (-2 + \pi \gamma\tau H_0(\gamma\tau)) + \gamma\tau J_0(\gamma\tau) (2 - \pi H_1(\gamma\tau)) \right)]^{-1} \\
    & \ \ \ \ \ \ \ \ \ \ \ \ \ \ \ \ \ \ \ \ \ \ \ \ \ \ \ \ \ \ \ \ \ \ \ \ \ \ \ \ \ \ \ \  \ \ \ \ \ \ \ \ \ \ \ \ \ \ \ \ \ \ \ \ \ \   \times\frac{2\beta\hbar\gamma}{\gamma^2\tau^2}\int^1_{0}d\omega \frac{ \ \text{Sin}^2(\gamma\tau\tilde{\omega}/2)}{\sqrt{\tilde{\omega}^2 - \tilde{\omega}^4}}\text{Coth}(\beta\hbar\gamma\tilde{\omega}/2), 
\end{align}
which cannot be calculated analytically. On the other hand, we can derive an analytic lower bound via the average zero-point energy,
\begin{align}\label{eq:boundszeropointenergy}
    \frac{1}{2}\hbar\beta\langle \omega \rangle_{\tilde{P}}=[\frac{\pi}{8 \gamma\tau} \left( J_1(\gamma\tau) (-2 + \pi \gamma\tau H_0(\gamma\tau)) + \gamma\tau J_0(\gamma\tau) (2 - \pi H_1(\gamma\tau)) \right)]^{-1}\times\frac{\beta\hbar\gamma}{8} \, _2F_3\left(1, 1; \frac{3}{2}, \frac{3}{2}, 2; -\frac{\gamma^2\tau^2}{4}\right),
\end{align}
where $_pF_q(a_1,..,a_p;b_1,...,b_q;x)$ is the generalised hypergeometric function. 
Fig.~\ref{fig:newexample} shows the behavior of the numerical Fano factor (solid lines) for this model as a function of the rescaled time $\gamma\tau$ and for different values of the temperature, together with the respective analytic lower bounds predicted by Eq. (17) in the main text (dashed lines). Overall, we see that our lower bound gives remains very tight at both short and long times for a wide range of temperatures.

\end{document}